\documentclass{elsart-3}

\usepackage{graphicx}

\usepackage{amssymb}
\usepackage{comment}

\usepackage[english,francais]{babel}

\newtheorem{e-proposition}[theorem]{Proposition}

\newtheorem{e-definition}[theorem]{Definition\rm}

\renewcommand{\theequation}{\arabic{equation}}
\def\og{\leavevmode\raise.3ex\hbox{$\scriptscriptstyle\langle\!\langle$~}}
\def\fg{\leavevmode\raise.3ex\hbox{~$\!\scriptscriptstyle\,\rangle\!\rangle$}}

\usepackage{color}
\definecolor{pink}{rgb}{1,1,0} 
\definecolor{red}{rgb}{1,0,0} 
\definecolor{blue}{rgb}{0,0,1} 
\definecolor{orange}{rgb}{1,0.5,0} 
\newcommand{\red}{\color{red}}

\newcommand{\gr}{\color{green}}
\newcommand{\blue}{\color{blue}}

\begin{document}

\begin{frontmatter}


\selectlanguage{english}
\title{Rarefied gas correction for the bubble entrapment singularity in drop impacts}


\selectlanguage{english}
\author[irphe]{Laurent Duchemin}
\author[dalembert]{and Christophe Josserand}

\address[irphe]{IRPHE, CNRS \& Aix-Marseille Universit\'e, 49 rue Joliot-Curie, 13013 Marseille, France}
\address[dalembert]{Institut D'Alembert, CNRS \& UPMC (Univ. Paris 6) UMR 7190,\\ 4 Place Jussieu, 75005 Paris, France}


\medskip
\begin{center}
{\small Received *****; accepted after revision +++++\\
Presented by Emmanuel Villermaux}
\end{center}

\begin{abstract}
We study the non-continuous correction in the dynamics of drop impact on a solid substrate. Close to impact, a thin film of gas is formed beneath the drop so that the local Knudsen number is of order one. We consider the first correction to the dynamics which consists of allowing slip of the gas along the substrate and the interface. We focus on the singular dynamics of entrapment that can be seen when surface tension and liquid viscosity can be neglected. There we show that different dynamical regimes are present 
that tend to lower the singularity strength. We finally suggest how these effects might be connected to the influence of the gas pressure in the impact dynamics observed in recent experiments.
{\it To cite this article: L. Duchemin and C. Josserand, C. R.
Mecanique 333 (2012).}

\vskip 0.5\baselineskip

\vskip 0.5\baselineskip
\noindent{\small{\it Mots-cl\'es~:} impact de gouttes; gaz rar\'efi\'es}}
\end{abstract}
\end{frontmatter}


\selectlanguage{english}
\section{Introduction}
\label{}
Drop impact is crucial in many multiphase flows ranging from raindrops to combustion chambers or ink-jet printing and it has become an emblematic problem of surface flows~\cite{Edge54,Rein93}. Depending on the impact parameters (drop diameter, velocity), fluid properties (viscosity, density, surface tension) and impacted surface (liquid deep or thin film, solid substrate), it can lead to many different outcomes: spreading, rebound, prompt splash, crown splash, cavity and jet formation to cite the most famous ones~\cite{Rio01}. Often, the influence of the surrounding gas is neglected in the analysis because of the high density and viscosity ratios between the gas and drop liquid. Indeed, beside the entrapment of a gas bubble at impact due to lubrication effect underneath the drop~\cite{thoroddsen03,thoroddsen05} and some aerodynamic corrections to the corolla dynamics, no significant effects of the gas was noticed so far. However, the situation has suddenly changed recently in a striking experiments on drop impacts on a smooth solid substrate~\cite{Xu05}: there, by changing only the operating pressure, they observe that splashes were suppressed as the pressure was lower below a critical level, emphasizing thus the crucial role played by the
gas in the splashing dynamics.
Although different {\it scenari} has been proposed to explain such effect, involving in particular gas compressibility~\cite{Xu05}, singular bubble entrapment dynamics~\cite{Mandre09,Mani10}, thin film skating~\cite{DJ11} and film wetting dynamics~\cite{kol12,DrNa11}, the surrounding gas influence remains yet a vibrant question of scientific debates.\\
Of particular interest is the coupled dynamics between the drop and the gas underneath it just before the impact. In this case, it can be shown that the thin film air dynamics can be considered within the lubrication approximation while the liquid viscosity can be neglected as far as thin liquid jets are not formed~\cite{Mandre09,DJ11,Korob08}. Then neglecting the surface tension one can see that a finite time singularity arises as a gas bubble is entrapped by the dynamics. This singularity behavior has to be regularized physically at least by the surface tension and the liquid viscosity but it has been argued that the resulting violent dynamics might be relevant in the splashing dynamics. Within the lubrication approximation for the gas film, a liquid jet is then formed that skates on the very thin (but non zero) gas layer. Eventually it has been shown experimentally that the liquid wets the solid substrate~\cite{kol12,DrNa11}, something that cannot be explained within such lubrication approximation when surface tension is present. In fact, different effects can be proposed to explain the liquid contact with the substrate when the gas layer is very thin: notably surface (Van der Waals for instance) interaction between the liquid interface and the substrate, interface fluctuations and/or surface roughness, and finite size (or rarefied gas) effect in the gas layer. In this paper, we focus on this latter case, that is when the gas layer thickness becomes of the order of the mean free path of the gas, leading firstly to a corrected lubrication equation for the thin film. In the next section, we recall the general scaling argument obtained for drop impact on a solid 
substrate using the classical lubrication equation for incompressible fluids. Then in section \ref{knudsen} we introduce the correction when the gas thickness is of the order of the mean-free path. Finally we discuss in section \ref{conclusion} the properties of the singularity in this case, in the absence of surface tension and we draw some perspectives for this work.

\section{Scaling analysis}
\label{scalings}

\begin{figure}
\begin{center}
\includegraphics[width=\linewidth]{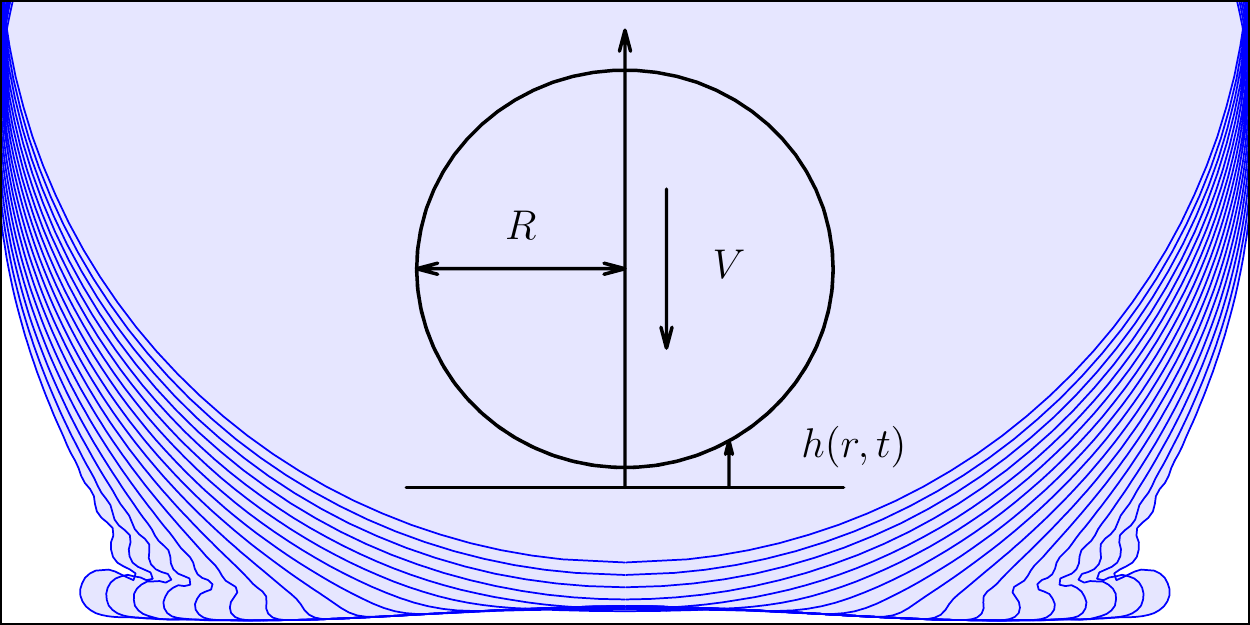}
\end{center}
\caption{\it{Sketch of the impacting drop and successive profiles 
for $We=\rho_l D U^2/\gamma=95$ and $St=1.35 \times 10^{-3}$.}} \label{sketch}
\end{figure}
We consider the impact of a liquid drop of diameter $D$ with a vertical velocity $U$ on a smooth solid substrate (Cf. Fig. \ref{sketch}). The liquid and surrounding gas densities are noted $\rho_l$ and $\rho_g$, their dynamical viscosity $\mu_l$ and $\mu_g$ respectively, the surface tension $\gamma$. Considering the high Reynolds and Froude numbers of the drop for typical experimental conditions
$${\rm Re}=\frac{\rho_l U D}{\mu_l} \;\;\; {\rm Bo}=  \frac{U}{\sqrt{gD}}$$
we assume that we can neglect the liquid viscosity and the gravity in the dynamics that is driven by the cushioning of the air film beneath the drop at impact.\\
Therefore, considering incompressible fluids we obtain the following set of equations describing the evolution of the drop surface $h(r,t)$ in axisymmetric geometry when it is approaching the substrate:
\begin{eqnarray}
(\partial \Omega) & \partial_t \varphi + \displaystyle \frac12 \nabla \varphi^2 +  \frac{p}{\rho_l} +\frac{\gamma}{\rho_l} \kappa= C(t), \label{eq:bern}\\
(\partial \Omega) & \partial_t h = \displaystyle \frac{1}{12 r \mu_g}\partial_r ( r h^3 \partial_r p ), \label{eq:lub}\\
(\partial \Omega) & \partial_t h = \partial_z \varphi  -\partial_r \varphi \partial_r h, \label{eq:inter}\\
(\Omega) & \Delta \varphi= 0, \label{eq:incom}
\end{eqnarray}
where $\varphi$ is the liquid velocity potential (the velocity in the liquid is ${\bf u}= {\bf \nabla} \varphi$), and $\Omega$ is the drop
volume, $\partial \Omega$ its interface.
The first equation (\ref{eq:bern}) is the Bernoulli equation valid in the liquid and written at the interface, while the last one 
(\ref{eq:incom}) is the incompressible condition for such potential flow. The interface motion is described by the advection equation (\ref{eq:inter}) while the lubrication equation (\ref{eq:lub}) allows the determination of the pressure $p(r,t)$ in the
gas film. The $1/12$ numerical prefactor in this equation was obtained by considering no slip condition for the gas velocity both on the solid substrate and on the fluid interface. This system of equations can be solved numerically using a boundary integral method and we resort with numerical simulation on the interface only~\cite{DJ11}.
Notice that lubrication is only valid where the slope of the interface is small enough and additional terms should be considered otherwise.\\
As the drop is approaching the solid substrate, cushioning of the gas leads to high pressure gradient beneath the drop. The high 
pressure created at the bottom of the drop deforms its shape so that a gas pocket is entrapped around the impact center. Simple
scaling arguments can help to estimate the typical sizes of this entrapped gas bubble. Indeed, considering the time $t=0$ as the
time of impact in the absence of air, one can estimate the typical vertical $H$ and horizontal $R$ scales of the drop deformation as $H(t)\sim Ut$ and $R(t) \sim \sqrt{DH}$ thanks to a geometrical argument based on the intersection between the falling drop and the substrate. Introducing these scalings in the lubrication equation yields the following scaling for the lubrication pressure in the gas layer:
$$ P_l \sim \frac{\mu_g R^2 U}{H^3}.$$
On the other hand, considering the impact pressure in the liquid needed to deviate horizontally a volume of liquid of typical size $R$ (this argument was first given in~\cite{JZ03}), one obtains~:
$$ P_i \sim \frac{1}{R^2} \frac{d}{dt} \left( \rho_l R^3 U \right) \sim \rho_l U^2 \frac{D}{R}. $$  
Thus, the lubrication pressure $P_l$ is strong enough to deviate the liquid until a critical height $H^*$ where the liquid somehow has to touch the solid substrate, that is when $P_i=P_l$, yielding:
$$ H^* \sim D \left(\frac{\mu_g}{\rho_l U D} \right)^{2/3}= D {\rm St}^{2/3}$$
introducing the Stokes number as the inverse of a Reynolds number balancing the liquid inertia with the gas viscous effects:
$$ {\rm St}= \frac{\mu_g}{\rho_l U D}.$$
Therefore, one expects the liquid to contact the substrate for $t^* \sim {\rm St}^{2/3} D/U$ entrapping a gas bubble of height $H^*$ and radius $R^* \sim D {\rm St}^{1/3}$. Notice that in practical (experimental) situations, the Stokes number is very small: for instance for a $2$ mm diameter droplet of water impacting at a velocity of $1$ meter per second, we obtain ${\rm St} \sim 10^{-8}$, so that the entrapped bubble radius is only a few thousands of the drop diameter.\\

However, it has been observed in numerical simulations of the set of equations (\ref{eq:bern},\ref{eq:lub},\ref{eq:inter},\ref{eq:incom}) that this contact would arise as a finite time singularity in the absence of surface tension (taking $\gamma=0$ in equation \ref{eq:bern})~\cite{Korob08,Mandre09,DJ11}. Such singularity exhibits a divergence of the pressure following $P \sim h_{min}^{-1/2}$ and of the interface curvature $\kappa \sim  h_{min}^{-2}$ which can be explained following~\cite{DJ11}.
Writing the set of governing equations in the frame moving radially with the geometrical intersection between the falling sphere and 
the substrate and using the dimensionless variables defined by
$$ \tilde{x}=\frac{r-R(t)}{D}, \;\; \tilde{z}=\frac{z}{D},\;\; \tilde{t}=\frac{Ut}{D},\;\; \tilde{h}(\tilde{x},\tilde{t})=\frac{h(r,t)}{D},\;\;
\tilde{p}=\frac{p}{\rho_l U^2},\;\; {\rm and} \;\; \tilde{\varphi}=\frac{\varphi}{UD},$$
the following system of equations is obtained:
\begin{eqnarray}
\partial_{\tilde{t}} \tilde{\varphi}-\dot{\tilde{R}} \partial_{\tilde{x}} \tilde{\varphi} + \displaystyle \frac12 \tilde{\nabla} \tilde{\varphi}^2 +  \tilde{p}= C(t), \label{eq:dimbern}\\
 \partial_{\tilde{t}} \tilde{h} -\dot{\tilde{R}} \partial_{\tilde{x}} \tilde{h}= \displaystyle \frac{1}{12 {\rm St} (\tilde{x} +{\rm St}^{1/3})}\partial_{\tilde{x}} \left((\tilde{x} +{\rm St}^{1/3})  \tilde{h}^3 \partial_{\tilde{x}} \tilde{p} \right), \label{eq:dimlub}\\
 \partial_{\tilde{t}} \tilde{h} -\dot{\tilde{R}} \partial_{\tilde{x}} \tilde{h}= \partial_z \varphi  -\partial_r \varphi \partial_r h, \label{eq:diminter}\\
\tilde{\Delta} \tilde{\varphi}= 0. \label{eq:dimincom}
\end{eqnarray}
The dimensionless radius $\tilde{R}=\sqrt{H/D}$ gives for the singularity a dimensionless radial velocity $\dot{\tilde{R}} \sim \sqrt{D/H}$.\\
Thus we develop the former set of equations near the singularity where $\dot{R} \sim {\rm St}^{-1/3}$ (we drop the $\tilde{}$ thereafter for the sake of simplicity) and assume that the time derivatives are subdominant compared with the $\dot{R} \partial_x$ terms, in good agreement with the numerics~\cite{DJ11}. Then seeking for a self-similar structure for the interface near the singularity in the form:
$$ h(x,t)=h_{min} f(\frac{x}{l(t)}), \;\; p(x,t)=P_0(t) g(\frac{x}{l(t)})\;\; {\rm and}\;\; \varphi(x,t)= \varphi_0(t) \Phi(\frac{x}{l(t)},\frac{z}{l(t)})$$
we obtain the following system of equations at the interface at the dominant order in the self similar variable $\xi=x/l(t)$ and $\chi=z/l(t)$:
\begin{eqnarray}
-{\rm St}^{-1/3} \frac{\varphi_0(t)}{l(t)} \partial_\xi \Phi+ \frac{\varphi_0(t)^2}{2 l(t)^2} \left( \nabla \Phi \right)^2+ P_0(t) g=C(t) \label{eq:selfbern}\\
-{\rm St}^{-1/3} \frac{h_{min}(t)}{l(t)} f'=\frac{1}{12 {\rm St}} \frac{h_{min}(t)^3 P_0(t)}{l(t)^2} (f^3g' )' \label{eq:selflub}\\
-{\rm St}^{-1/3} \frac{h_{min}(t)}{l(t)} f'=\frac{\varphi_0(t)}{l(t)} \left(\partial_\chi \Phi- \frac{h_{min}(t)}{l(t)} \partial_\xi \Phi f' \right) \label{eq:selfadv}
\end{eqnarray}

where the prime stands for the derivative of the function to the variable $\xi$. The lubrication equation (\ref{eq:selflub}) gives the following relation:
$$ P_0(t) \sim {\rm St }^{2/3} \frac{l(t)}{h_{min}(t)^2},$$
and from the interface dynamics eq. (\ref{eq:selfadv}) one can see that two situations have to be considered:
\begin{itemize}
\item $h_{min}(t) \ll l(t)$ leading to the observed numerical scalings since $\varphi_0(t)\sim {\rm St }^{-1/3}h_{min}(t)$ that gives $P_0(t) \sim {\rm St }^{-2/3} h_{min}(t)/l(t)$ and thus:
$$ l(t) \sim {\rm St }^{-2/3} h_{min}(t)^{3/2},\;\; P_0(t) \sim h_{min}(t)^{-1/2} \;\;{\rm and} \;\; \kappa \sim {\rm St}^{4/3} h_{min}(t)^{-2}.$$
and this regime is found to be valid for {\it thick} films, $h_{min} \gg {\rm St}^{4/3}$.

\item $h_{min}(t) \gg l(t)$, which gives $\varphi_0(t)\sim {\rm St }^{-1/3}l(t)$ and yielding:
$$ P_0(t) \sim {\rm St}^{-2/3},\;\; l(t) \sim {\rm St }^{-4/3} h_{min}(t)^2 \;\; {\rm and} \;\; \kappa \sim {\rm St}^{8/3} h_{min}(t)^{-3}.$$
Similarly, this regime is valid for {\it thin} films, $h_{min} \ll {\rm St}^{4/3}$. Finally, notice that this second regime has never been reached numerically so far because of the small values of the Stokes number considered.
\end{itemize}

\section{Influence on the mean free path}
\label{knudsen}
It is interesting to observe that under typical experimental conditions the Stokes number is very small so that the system of equations used has to be questionned. In particular, when the typical air layer becomes of the order of the mean free path $\lambda$ one has to consider rarefied gas correction to the gas dynamics. This can be quantified by the Knudsen number ${\rm Kn}$ defined as the ratio between the mean free path and the typical air layer thickness which gives for the entrapped bubble described above:
$$ {\rm Kn} = \frac{\lambda}{D {\rm St}^{2/3}},$$
and one expects rarefied gas effects to enter into account for ${\rm Kn} \gg 0.01$. 
From kinetic theory, we have that the mean free path is related to the gas pressure $P_g$ through:
$$ \lambda = \frac{k_B T}{\sqrt{2}\pi d^2 P_g},$$
where $T$ is the temperature, $k_B$ the Boltzmann constant and $d$ the typical size of the atoms of the gas. For ambient 
temperature $T=300$ K and ambient pressure $P_{g0}=10^5$ Pa the typical mean free path in the air is 
$ \lambda_0 =70 \, {\rm nm}$ and one can write the simple relation:
$$ \frac{\lambda}{\lambda_0}= \frac{P_{g0}}{P_g}.$$
In the situation of incompressible fluids (gas and liquid) considered here, $\lambda$ is a constant that is only a function of the 
ambient pressure and it does not formally depends on the dynamical pressure used in the equations. In fact, such approximation
is valid as long as one can neglect the gas compressibility which should be accounted for otherwise. The effect of the gas 
compressibility has been studied in details in~\cite{Mandre09,Mani10,Mandre12} using a low Mach number lubrication approximation but no crystal clear 
mechanism involving directly the gas compressibility could be identified in the drop splashing, beside some complicated 
dependance of the dynamics. Therefore, in order to disentangle the influence of the rarefied gas correction from the compressible
influence, we keep the incompressible limit in the dynamical equation, considering only the influence of the Knudsen number in 
the boundary conditions as explained below. Finally, as the rarefied gas situation is intrinsically for compressible fluids, an 
extension of this work might have to consider compressible effects.

Since $\lambda$, and thus ${\rm Kn}$, increases as the gas pressure decreases, one can expect that the correction due to the mean free path of the gas has to be considered when the gas pressure is lowered. Indeed, taking the experimental
conditions of~\cite{Xu05} where the striking influence of the air pressure was first illustrated, one finds for atmospheric
pressure ${\rm Kn}=0.66$ so that correction due to the rarefied gas configuration has to be investigated.\\
For such relatively high Knudsen numbers, a simple way to account for the correction due to the rarefied gas context is to introduce a slipping velocity for the gas so that the no-slip boundary condition at a solid interface transforms into the Navier-slip condition:
$$ u_t= \lambda \frac{\partial u_t}{\partial n},$$
where $n$ is the normal direction at the interface and $u_t$ the tangential velocity. This condition comes from the fact that at the level of the mean free path the no-slip condition is meaningless and cannot be imposed. Solving the Stokes equation between the solid substrate and the drop interface located at $z=h(r,t)$ and assuming that the Navier-slip condition applies on both sides we obtain the following relation for the radial velocity $u$ in the gas layer, under the thin film approximation where the interface slope is supposed to be small:

\begin{equation} 
u(z,t)=-\frac{1}{2 \mu_g} \partial_r p \left(hz-z^2+\lambda h \right),
\label{u_lub}
\end{equation}
so that the lubrication equation (\ref{eq:lub}) becomes:
\begin{equation}
\partial_t h = \displaystyle \frac{1}{12 r \mu_g}\left( \partial_r ( r h^3 \partial_r p )+6\lambda  \partial_r ( r h^2 \partial_r p ) \right).
\label{eq:rare}
\end{equation}
Therefore, for film thickness $h \ll \lambda$ one expects that the second term in the lubrication dominates so that the singularity
features should be changed. 

\section{Scaling in the rarefied gas limit}
\label{rarefied}
The general dynamics of air cushioning during drop impact will thus be governed by equation (\ref{eq:rare}) and different dynamical regimes will dominate depending on the air layer thickness. When the air layer is always much thicker than the mean 
free path $\lambda$ then the former equation (\ref{eq:lub}) will be valid and the scaling obtained in section \ref{scalings} will be
observed. On the other hand, when $h \ll \lambda$ the lubrication dynamics will be dominated by the new term introduced in (\ref{eq:rare}) so that the following equation should be investigated:

\begin{equation}
\partial_t h = \displaystyle \frac{\lambda }{2 r \mu_g}\left( \partial_r ( r h^2 \partial_r p ) \right),
\label{eq:slip}
\end{equation}
where the usual lubrication term is neglected. In between and when the air layer is varying within these limits, one has to 
study the general equation (\ref{eq:rare}). Moreover, two different situations can be identified: firstly, 
$Kn \ll 1$ so that the bubble entrapment still arises around the same radius $D {\rm St}^{1/3}$ than above, although the dynamics of the film cushioning at the singularity neck will eventually be governed by the slipping equation (\ref{eq:slip}). On the other hand, 
for  $Kn >1$, already the dynamics of the bubble entrapment has to be considered within the rarefied gas limit of eq. (\ref{eq:slip}) 
so that the radius of the bubble itself (and thus the radial velocity of the interface near the singularity) is changed.

\subsection{Low Knudsen numbers case}
First of all, we thus consider that the gas bubble entrapped has the same features than before. This assumption is reasonable for small enough Knudsen number and the bubble formation is dominated by the usual lubrication equation. However, the 
singularity dynamics itself has to be studied within the rarefied gas limit since the local air layer becomes much thinner than the 
mean free path. Then we can write the dynamics in the frame moving with the former singularity velocity $U {\rm St}^{-1/3}$. Using the same dimensionless units and then seeking for similarity solution using the same change of variables than before, 
we obtain at the dominant order near the singularity:

\begin{eqnarray}
-{\rm St}^{-1/3} \frac{\varphi_0(t)}{l(t)} \partial_\xi \Phi+ \frac{\varphi_0(t)^2}{2 l(t)^2} \left( \nabla \Phi \right)^2+ P_0(t) g=C(t) \label{eq:rarebern}\\
-{\rm St}^{-1/3} \frac{h_{min}(t)}{l(t)} f'=\frac{1}{12 {\rm St}} \frac{h_{min}(t)^3 P_0(t)}{l(t)^2} \left(f^3g' +6\frac{\lambda}{h_{min}(t)} f^2 g' \right)' \label{eq:rarelub}\\
-{\rm St}^{-1/3} \frac{h_{min}(t)}{l(t)} f'=\frac{\varphi_0(t)}{l(t)} \left(\partial_\chi \Phi- \frac{h_{min}(t)}{l(t)} \partial_\xi \Phi f' \right) \label{eq:rareadv}
\end{eqnarray}

Notice that $\lambda$ is now the mean free path made dimensionless using the drop diameter. Now, two regimes can also be identified due to the two terms in the right hand side of the lubrication equation (\ref{eq:rarelub}).

\begin{itemize}
\item $h_{min}(t) \gg \lambda$, the pressure relation due to the dominant term in the lubrication equation remains:
$$ P_0(t) \sim {\rm St }^{2/3} \frac{l(t)}{h_{min}(t)^2}.$$

\item $h_{min}(t) \ll \lambda$, the pressure relation is determined by the other term in the lubrication equation, yielding
$$ P_0(t) \sim {\rm St }^{2/3} \frac{l(t)}{\lambda h_{min}(t)}.$$
\end{itemize}

Finally, we resort with two dynamical {\it scenari} for the singularity dynamics, depending on the ratio between the mean free path
$\lambda$ and the critical thickness ${\rm St}^{4/3}$ separating the two regimes of the singularity without slip condition, namely:

\begin{itemize}
\item $\lambda \gg {\rm St}^{4/3}$, two dynamical regimes follow. First, when $h_{min} \gg \lambda$, then the "usual" scaling are 
valid:
$$ l(t) \sim {\rm St }^{-2/3} h_{min}(t)^{3/2},\;\; P_0(t) \sim h_{min}(t)^{-1/2} \;\;{\rm and} \;\; \kappa \sim {\rm St}^{4/3} h_{min}(t)^{-2}.$$
then it is followed by another regime as $h_{min}$ decreases. When $h_{min} \ll \lambda$, 
the following scalings are obtained: 
$$ l(t) \sim \sqrt{\lambda} h_{min}(t) {\rm St}^{-2/3},\;\; P_0(t) \sim \frac{1}{\sqrt{\lambda}} \;\; {\rm and} \;\; \kappa \sim \frac{{\rm St}^{4/3}}{\lambda h_{min}(t)}.$$
and we remain thereafter within the configuration where $h_{min}(t)/l(t) \sim {\rm St}^{2/3}/\sqrt{\lambda} \ll 1$. In this case, the regime where $h_{min}(t) \gg l(t)$ is not present.

\item  $\lambda \ll {\rm St}^{4/3}$, then for $h_{min} \gg {\rm St}^{4/3} \gg \lambda$ the "usual" scaling holds:
$$ l(t) \sim {\rm St }^{-2/3} h_{min}(t)^{3/2},\;\; P_0(t) \sim h_{min}(t)^{-1/2} \;\;{\rm and} \;\; \kappa \sim {\rm St}^{4/3} h_{min}(t)^{-2}.$$
It is followed by a regime where $ {\rm St}^{4/3} \gg h_{min} \gg \lambda$ which gives the second scaling obtained in the beginning
$$ P_0(t) \sim {\rm St}^{-2/3},\;\; l(t) \sim {\rm St }^{-4/3} h_{min}(t)^2 \;\; {\rm and} \;\; \kappa \sim {\rm St}^{8/3} h_{min}(t)^{-3}.$$
Such regime is also followed by a new regime when $ {\rm St}^{4/3} \gg \lambda  \gg h_{min}$ yielding:
$$ l(t) \sim \lambda h_{min}(t) {\rm St}^{-4/3} \;\; P_0(t) \sim {\rm St}^{-2/3} \;\; {\rm and} \;\; \kappa \sim \frac{{\rm St}^{8/3}}{ \lambda^2 h_{min}(t)}.$$

\end{itemize}

\subsection{Full rarefied gas limit}
In the experiments described in~\cite{Xu05}, the Knudsen number based on the typical 
bubble thickness obtained with the usual lubrication equation is of order one and it increases when the pressure is lowered. 
Therefore, the rarefied gas effect has to be accounted for in the dynamics of the bubble 
entrapment and the above scalings fail already for the size of the entrapped bubble. 
In this regime, we have $\lambda \gg D {\rm St}^{2/3} \gg h(r,t)$ everywhere and the second term in the lubrication equation -- proportional to $\lambda$ --
is dominant. Therefore, the scaling for the lubrication pressure reads~:
$$
P_l \sim \frac{\mu_g R^2 U}{\lambda H^2}
$$
Balancing the impact pressure with this lubrication pressure, 
one obtains for the typical bubble height~:
$$
H^* \sim D \frac{St^2 D^2}{\lambda^2}.
$$
It needs to be emphasized that we obtain here a bubble height $H^*$ and radius 
$$ R^* =\sqrt{DH} \sim   D {\rm St} \frac{D}{\lambda},$$
that depend on the external gas pressure through $\lambda$! Notably, the bubble radius 
is proportional to the external pressure and therefore decreases when this ambient  pressure 
is lowered.
Considering again that the singularity follows the geometrical radial scale $R\sim\sqrt{DUt}$, 
we obtain that the radial velocity of the singularity yields:
$$
\dot{R^*}\sim U \frac{\lambda}{ D {\rm St}},
$$

Again, we can assume that the time derivatives are subdominant compared to the $\dot{R} \partial_x$ terms.
Then, after writing the dynamical equation in dimensionless form and seeking for self-similar 
solutions, we obtain the following system of equation at the interface at the dominant order in the self similar variable $\xi=x/l(t)$ and $\chi=z/l(t)$:

\begin{eqnarray}
-\frac{\lambda}{{\rm St}} \frac{\varphi_0(t)}{l(t)} \partial_\xi \Phi+ \frac{\varphi_0(t)^2}{2 l(t)^2} \left( \nabla \Phi \right)^2+ P_0(t) g=C(t) \label{eq:knudbern}\\
-\frac{\lambda}{{\rm St}} \frac{h_{min}(t)}{l(t)} f'=\frac{1}{2 {\rm St}} \frac{h_{min}(t)^2 \lambda P_0(t)}{l(t)^2} \left( f^2 g' \right)' \label{eq:knudlub}\\
-\frac{\lambda}{{\rm St}} \frac{h_{min}(t)}{l(t)} f'=\frac{\varphi_0(t)}{l(t)} \left(\partial_\chi \Phi- \frac{h_{min}(t)}{l(t)} \partial_\xi \Phi f' \right) \label{eq:knudadv},
\end{eqnarray}
where we have assumed that $\lambda \gg h$ is true everywhere in the bubble region so that only the slipping term in equation (\ref{eq:knudlub}) is present.

Therefore, balancing the two terms in the lubrication equation, we obtain for the pressure~:
$$
P_0(t) \sim \frac{l(t)}{h_{min}(t)}.
$$

Then, we get again {\it a priori} two different regimes, depending on the value of $h_{min}$ compared to the value of $l(t)$~:

\begin{itemize}
\item $h_{min}(t) \ll l(t)$, leading to $\varphi_0(t)\sim \lambda h_{min}(t) / {\rm St }$ that gives~:
$$ l(t) \sim \lambda h_{min}(t) {\rm St }^{-1},\;\; P_0(t) \sim \lambda {\rm St }^{-1} \;\;{\rm and} \;\; \kappa \sim {\rm St}^{2} \lambda^{-2} h_{min}(t)^{-1},$$
and this regime is found to be valid when $\lambda \gg {\rm St}$, which is always the case since this regime was found to be valid for $\lambda \gg {\rm St}^{2/3} \gg {\rm St}$ (since ${\rm St} \ll 1$ 
in the experiments).

\item $h_{min}(t) \gg l(t)$, would give $\varphi_0(t)\sim \lambda l(t) / {\rm St }$ and yields:
$$ l(t) \sim \lambda^2 h_{min}(t) {\rm St }^{-2},\;\; P_0(t) \sim \lambda^2 {\rm St }^{-2} \;\;{\rm and} \;\; \kappa \sim {\rm St}^{4} \lambda^{-4} h_{min}(t)^{-1},$$
As explained above, this regime cannot be observed in realistic experiments, since it implies $\lambda \ll {\rm St}$, and 
therefore that the bubble height is much bigger that the radius of the drop. 
\end{itemize}

\newpage

\section{Discussion}
\label{conclusion}
We have obtained the correction due to the account of the rarefied gas limit in the singularity dynamics leading to a bubble 
entrapment during
drop impact. In particular, we have shown that the pertinent regime for the experiments, where the splashing properties are influenced by the gas pressure, corresponds to a situation where the bubble size reduces with the gas pressure. The question of vibrant scientific debates is whether and how one can relate this singularity to the splashing dynamics of
drop impact on solid substrate. Indeed, it has been often argued that such violent singular behavior is at the heart of the 
onset of splashing so that the regularization of the singularity would be crucial to determine the splashing properties~\cite{Mandre09,Mani10,DJ11,Mandre12}. In particular, surface tension on one side and liquid viscosity on the other side will both regularize the singularity, leading to a rapid jet propagating along the solid wall~\cite{DJ11}. However, such effects could not explain the experimental results of~\cite{Xu05} if only the usual 
(without the rarefied gas correction) incompressible lubrication dynamics was considered for the air layer. This is why compressible effects have been often investigated, 
in the framework of the low Mach number compressible lubrication equation~\cite{Mandre09,Mani10}. 
Finally, while within the lubrication 
approximation, the jet skates on the thin gas layer, it has to be noticed that experimentally the liquid eventually wets the
solid substrate~\cite{DrNa11,kol12} through a dynamical rupture of the gas layer that still needs to be elucidated.
Here we would like to emphasize that the rarefied gas effects accounted in the incompressible limit offers an alternative physical mechanism to explain the air pressure effect in the drop 
impact, as suggested also in the compressible model~\cite{Mandre12}. In particular, we have demonstrated here that the
entrapped bubble size depends strongly on the external gas pressure, a result that has to be investigated in future works since it changes the jet features and consequently the air drainage dynamics.

\section*{Acknowledgements}
It is our pleasure to thank Paul Clavin for stimulating discussions on this subject initiated in a workshop organized by him in Peyresq in 2009. Notably, we previously tackled a singularity in self-similar variables using boundary integral methods (in 2D then) in our first collaboration with Paul on the Rayleigh-Taylor non-linear asymptotic dynamics!\\

Finally, we would like to thank C\'edric Croizet and Yves Pomeau for interesting discussions on the rarefied gas dynamics.
C.J. wants to acknowledge the support of the Agence Nationale de la Recherche through its Grant ÒDEFORMATIONÓ ANR-09-JCJC-0022-01 and of the Programme \'Emergence(s) of the Ville de Paris.




\bibliography{knudsen}
\bibliographystyle{unsrt}
\end{document}

 In fact, because of the high dependance of the pressure with the gas layer thickness the pressure converges very rapidly to the constant external pressure so that a free surface condition ($p=P_0$) has to be applied in Bernoulli equation. In our numerical scheme lubrication equation is written in terms of the arc-length along the interface, 
then it is solved from the point beneath the drop located on the axis of symmetry to the first point where the interface is vertical (at short-time, 
this point is the equator of the drop; once a jet is formed, it is the radial extension of the jet). 
After this point, free surface condition is applied (see~\cite{DJ11} for more details).

\selectlanguage{francais}
\noindent{\bf R\'esum\'e}
\vskip 0.5\baselineskip
\noindent
{\bf  Dynamique singulire d'emprisonnement de bulle lors de l'impact d'une goutte \`a nombre de Knudsen \'elev\'e.}
Nous \'etudions la correction due \`a la limite des gaz rar\'efi\'es rencontr\' ee lors de l'impact de gouttes sur substrat solide. Cette
situation appara\^{\i}t sous la goutte dans le film d'air fin form\'e lors de l'impact. Nous prenons en compte la premi\`ere correction \`a la th\'eorie classique de lubrification lorsque le nombre de Knudsen devient d'ordre un, qui consiste \`a autoriser une vitesse de glissement \`a l'interface et au niveau du solide.
{\it Pour citer cet article~: L. Duchemin and C. Josserand, C. R.
Mecanique 333 (2012).}

